\documentclass{jetpl}
\twocolumn
\usepackage{graphicx}
\usepackage{bm}
\lat
\usepackage[usenames,dvipsnames]{color} 
\newcommand{\mkrm}[1]{}           

\title{Phonon-particle coupling effects in odd-even double mass differences of semi-magic nuclei}%

\rtitle{Phonon-particle coupling effects in odd-even double mass differences of semi-magic nuclei}%

\sodtitle{Phonon-particle coupling effects in single-particle energies of semi-magic nuclei.}%

\author{
 E.\,E.\,Saperstein$^{*,**}$\/\thanks{e-mail: saper43\_7@mail.ru},
 M.\,Baldo$^{\dagger}$,
 S.\,S.\,Pankratov$^{*,***}$,
 S.\,V.\,Tolokonnikov$^{*,***}$}

\rauthor{E.\,E.\,Saperstein, M.\,Baldo, S.\,S.\,Pankratov, S.\,V.\,Tolokonnikov}

\sodauthor{E.\,E.\,Saperstein, M.\,Baldo, S.\,S.\,Pankratov, S.\,V.\,Tolokonnikov}

\address{$^{*}$National Research Centre Kurchatov Institute, pl.
Akademika Kurchatova 1, Moscow, 123182 Russia\\
$^{**}$National Research Nuclear University MEPhI, 115409 Moscow, Russia\\
$^{\dagger}$Istitute Nazionale di Fisica Nucleare, Sezione di Catania, 64 Via S.-Sofia, I-95125 Catania, Italy\\
$^{***}$Moscow Institute of Physics and Technology, 141700 Dolgoprudny, Russia}

\PACS{21.60.-n; 21.65.+f; 26.60.+c; 97.60.Jd}

\dates{\today}{*}

\abstract{ A method is developed to consider the particle-phonon coupling (PC) effects in the
calculation of the odd-even double mass differences (DMD) in semi-magic nuclei  starting from the free
$NN$-potential. The PC correction $\delta \Sigma^{\rm PC}$ to the mass operator $\Sigma$ is found in
$g_L^2$-approximation, $g_L$ being the vertex of creating the $L$-phonon. The tadpole term of the
operator $\delta \Sigma^{\rm PC}$ is taken into account. The method is based on a direct, without any
use of the perturbation theory, solution of the Dyson equation with the mass operator
$\Sigma(\eps){=}\Sigma_0{+}\delta \Sigma^{\rm PC}(\eps)$ for finding the single-particle energies and
$Z$-factors. In its turn, they are used as an input for finding different PC corrections to the DMD
values. Results for a chain of even semi-magic nuclei $^{200-206}$Pb show that the inclusion of the PC
corrections makes agreement with the experimental data significantly better.}

\begin{document}

\newcommand{\beq}{\begin{equation}}
\newcommand{\eeq}{\end{equation}}
\newcommand{\bea}{\begin{eqnarray}}
\newcommand{\eea}{\end{eqnarray}}
\newcommand{\eps}{\varepsilon}
\newcommand{\bfg}{\boldsymbol}

\maketitle

Recently, we developed a method to find the particle-phonon coupling (PC) corrections to the odd-even
double mass differences (DMD) of magic nuclei starting from  the free $NN$-potential (FP)
\cite{DMD1,DMD2}. This problem is closely related to that of finding the pairing gap $\Delta$ in
nuclei in terms of the FP, which was rather popular in the last decade, see the review article
\cite{BCS50} and Refs. therein. The approach was essentially based on the so-called
$g_L^2$-approximation, where $g_L$ is the vertex of the $L$-phonon creation, which works well in magic
nuclei. The use  the perturbation theory (PT) in the PC correction $\delta \Sigma^{\rm PC}(\eps)$ with
respect to the initial mass operator $\Sigma_0$ for finding the  single-particle (SP) energies is an
additional approximation used in \cite{DMD1,DMD2}. In semi-magic nuclei, the latter is regularly not
valid due to presence of the low-lying collective $2^+$ phonon. In its turn, it results in the
appearance of small energy denominators in the expression for the operator $\delta \Sigma^{\rm
PC}(\eps)$ and in the one for the phonon induced interaction ${\cal V}_{\rm ind}$ as well, making the
use of the PT inapplicable. Recently the problem for SP energies was resolved by us in \cite{SPEnPB}
by direct solving the Dyson equation with the mass operator $\Sigma(\eps){=}\Sigma_0{+}\delta
\Sigma^{\rm PC}(\eps)$ without any use of the PT. In this work, we continue development of this method
to find the DMD values for semi-magic nuclei.

As in \cite{SPEnPB}, we deal with a normal sub-system of the nucleus under consideration, therefore
the general set of equations for DMD in magic nuclei \cite{DMD1,DMD2} remains valid. However, the
method of their solution for semi-magic nuclei changes significantly. We limit ourselves with Pb
isotopes, thus, the proton DMD values will be considered only which are defined in terms of the
nuclear masses $M(N,Z)$, where $N$ and $Z$ are neutron  and proton  numbers correspondingly in the
nucleus under consideration, as follows: \beq D_{2p}^+(N,Z) {=}  M(N,Z{+}2){+}
M(N,Z){-}2M(N,Z{+}1),\label{d2ppl} \eeq \beq D_{2p}^-(N,Z) {=} {-}M(N,Z-2){-}M(N,Z){+}
2M(N,Z{-}1).\label{d2pmi}\eeq

To make the discussion more transparent, we repeat schematically the main relations for DMD of semi-magic nuclei without PC corrections \cite{Gnezd-1,Gnezd-2}.  Let us start  from the Lehmann expansion  for the two-particle Green function $K$.  In the SP wave functions $|1\rangle{=}|n_1,l_1,j_1,m_1\rangle$ representation, it  reads \cite{AB}: \beq K_{12}^{34}(E)=\sum_s \frac {\chi^s_{12}\chi^{s+}_{34}} {E-E_s^{+,-} \pm i\gamma}, \label{Lem}\eeq where $E$ is the total energy in the two-particle channel and $E_s^{+,-}$ denote the energies of eigenstates of nuclei with two particles or two holes, respectively, added to the original nucleus. The lowest ones of them determine the mass differences entering Eqs. (\ref{d2ppl}) and (\ref{d2pmi}).

The Green function $K$ relates to the two-particle interaction amplitude $\Gamma$ as follows: \beq K = K_0 + K_0 \Gamma K_0, \label{gam}\eeq where $K_0=GG$, $G$ being the one-particle Green function. Within the Brueckner theory, the amplitude $\Gamma$ obeys the Bethe--Goldstone equation: \beq \Gamma = {\cal V}+{\cal V} GG \Gamma, \label{eqgam}\eeq where ${\cal V}$ is the FP.

This equation is in many ways similar to the Brueckner theory gap equation \cite{BCS50} possessing the same problem of slow convergence. For the Argonne v$_{18}$ potential we use as the FP, SP states with energies up to 1 GeV should be included into the SP space to obtain a good accuracy. Therefore, the same two-step renormalization method was used in \cite{Gnezd-1,Gnezd-2}
which was developed previously for the pairing problem in Refs. \cite{Pankrat-1,Pankrat-2}.   The complete two-particle Hilbert space $S$ of the problem is split in the model subspace $S_0$, including the SP states with energies less than a separation energy $E_0$, and the complementary one, $S'$. In practice, in all the articles cited above we use $E_0{=}40\;$MeV. In the result, Eq. (\ref{eqgam}) is split into two ones: in the model space, \beq \Gamma = {\cal V}_{\rm eff}+{\cal V}_{\rm eff} GG \Gamma|_{S_0}, \label{eqgam0}\eeq  and in the subsidiary space:\beq {\cal V}_{\rm eff} = {\cal V}  + {\cal V}  G  G {\cal V}_{\rm eff}|_{S'}. \label{Vef} \eeq To solve the last equation for the effective interaction (EI) ${\cal V}_{\rm eff}$, a method of ``Local Potential Approximation'' was developed for the pairing problem \cite{Pankrat-1,Pankrat-2}, with  the use of plane waves instead of the exact SP states $|\lambda\rangle$. In the DMD problem, \cite{Gnezd-1,Gnezd-2}, it turned out to be also applicable. As to the first of these two equations, it was solved in the space $S_0$ directly, without additional approximations. In this case, it is convenient to carry out the integration in Eq. (\ref{eqgam0}) of the product $GG$ of two Green functions over the relative energy: \beq\begin{aligned}A_{12}(E) =  \int &\frac {d\eps}{2\pi i}G_1\left(\frac E 2 {+}\eps \right) G_2\left(\frac E 2 {-}\eps \right) = \\ &= \frac {1{-}n_1{-}n_2} {E{-}\eps_1{-}\eps_2}\,,\end{aligned}\label{Alam}\eeq where $\eps_{1,2}$ are the SP energies and $n_{1,2}{=}(0;1)$, the corresponding occupation numbers.

In Refs. \cite{Pankrat-1,Pankrat-2}, the ``semi-microscopic model'' was suggested to take into account approximately many-body theory corrections to the EI (\ref{Vef}) found in terms of the FP. The main term (\ref{Vef})  is supplemented with a phenomenological
$\delta$-function addendum: \beq\begin{aligned}{\cal V}_{\rm eff}({\bf r}_1,...,{\bf r}_4) &= V^{\rm free}_{\rm eff}({\bf r}_1,...,{\bf r}_4) + \\ + &\gamma C_0 \frac {\rho(r_1)}{\bar{\rho}(0)} \prod_{k=2}^4\delta ({\bf r}_1 {-} {\bf r}_k)\,.\end{aligned}\label{Vef1}\eeq Here $\rho(r)$ is the density of nucleons of the kind under consideration (protons in our case), $C_0{=}300\, {\rm MeV} {\cdot}\, {\rm fm}^3$ is the usual normalization factor of the theory of finite Fermi systems \cite{AB}, and $\gamma$ is a dimensionless phenomenological parameter. The quantity ${\bar{\rho}(0)}$ in the denominator is the average central density. The value of $\gamma{=}0.06$ was found in the references above as an optimal one for describing the bulk of data on the pairing gap. It turned out to be successful also for describing the DMD values in magic and semi-magic nuclei \cite{Gnezd-1,Gnezd-2} without PC corrections. In Refs. \cite{DMD1,DMD2}, it was shown that, after inclusion of the PC corrections in magic nuclei, this addendum is diminished to  $\gamma{=}0\div  0.03$. Here, a similar analysis is carried out for semi-magic nuclei.

The DMD values without PC corrections are determined with the eigenenergies $E_s$ of the following equation \cite{Gnezd-1}: \beq (E_s-\eps_1-\eps_2) \chi^s_{12}=(1-n_1-n_2) \sum_{34\in S_0} ({\cal V}_{\rm eff})_{12}^{34} \chi^s_{34}\label{eqchi}.\eeq It is different from the Shr\"{o}dinger equation for two interacting particles in an external potential well only by the factor $(1{-}n_1{-}n_2)$ reflecting the Pauli principle. Just as in the pairing problem, the angular momenta of two-particle states  $|12\rangle$, $|34\rangle$ are coupled to the total angular momentum $I{=}0$ ($S{=}0, L{=}0$).

To include the low-lying phonons, we should take into account that they influence mainly the SP states close to the Fermi level. Therefore, it is reasonable to make an additional renormalization of Eq. (\ref{eqchi}) by splitting our model space $S_0$ to the ``valence'' subspace $S_0^{\rm v}$, containing two shells adjacent to the Fermi level, and the subsidiary part $S'_0$ of the model space.

\begin{figure}
\centerline {\includegraphics [width=40mm]{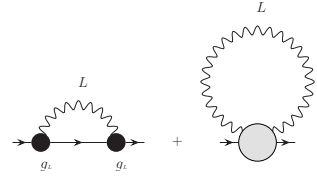}} \caption{Fig.~\ref{fig:SigPC}. PC corrections to the mass operator. The black circle is the vertex $g_L$ of creating the $L$-phonon, the gray blob denotes the phonon ``tadpole'' term.} \label{fig:SigPC}
\end{figure}

To find the SP energies  in the space $S_0^{\rm v}$ with account for the PC effects, we solve the following equation: \beq \left(\eps-H_0 -\delta \Sigma^{\rm PC}(\eps) \right) \phi =0, \label{sp-eq}\eeq where $H_0$ is the quasiparticle Hamiltonian with the spectrum $\eps_{\lambda}$, and $\delta \Sigma^{\rm PC}$ is the PC correction to the quasiparticle mass operator. All diagrams for it of the order $g_L^2$ are displayed in Fig.~1. If the PT  in $\delta \Sigma^{\rm PC}$ with
respect to $H_0$  is valid, as it occurs in magic nuclei, we have for the PC corrected SP energies \cite{DMD1,DMD2,levels}: \beq\begin{aligned}\eps_{\lambda} &= \eps_{\lambda} + Z_{\lambda}\delta \Sigma^{\rm PC}_{\lambda\lambda}(\eps_{\lambda})\,, \\ Z_{\lambda} &= \left( 1 - \left(\frac {\partial} {\partial \eps} \delta \Sigma^{\rm PC}(\eps) \right)_{\eps=\eps_{\lambda}}\right)^{-1}\,.\end{aligned}\label{eps-Z-fac}\eeq In this case, each SP state in the valent space generates the single PC corrected one: $|\lambda\rangle \to \tilde{|\lambda\rangle}{=} \sqrt{Z_{\lambda}}|\lambda\rangle$, and we obtain from (\ref{eqchi}) the PC corrected equation in the valence space:
\beq (E_s-\tilde \eps_1-\tilde \eps_2) \chi^s_{12}=(1-n_1-n_2) \sum_{34\in S_0^{\rm v}} (\tilde {\cal V}_{\rm eff})_{12}^{34} \chi^s_{34}\label{eqchi0},\eeq
\beq \langle 1 1'|\widetilde{\cal V}_{\rm eff}|22'\rangle
 =  \sqrt{Z_1 Z_{1'} Z_2  Z_{2'} }
\langle 1 1'|{\cal V}_{\rm eff}^{\rm v} +  {\cal V}_{\rm ind}|22'\rangle, \label{Vtild}\eeq
where the  EI ${\cal V}_{\rm eff}^{\rm v}$ obeys the equation similar to (\ref{eqgam}), but in the subsidiary model space $S'_0$. This equation is solved directly, without any approximation. The phonon induced interaction ${\cal V}_{\rm ind}$ is displayed in Fig.~2. The explicit expression of the matrix element of ${\cal V}_{\rm ind}$ is as follows \cite{DMD1,DMD2}: \bea \langle 1 1'|{\cal V}_{\rm ind}|22'\rangle = - \frac {2 \omega_L} {\sqrt{(2j_1+1)(2j_2+1)}}  \nonumber \\
\times \frac {\bigl(\langle j_1 l_1|| Y_L || j_1 l_1\rangle (g_L)_{11'}\bigr) \bigl(\langle j_2 l_2|| Y_L || j_2 l_2\rangle (g_L)_{22'}\bigr)^* }{\omega_L^2-(\eps_2-\eps_1)^2 }, \label{Vind} \eea where $\omega_L$ is the excitation energy of the $L$-phonon,
$\langle \;|| Y_L || \;\rangle$ stands for the reduced matrix element, and $(g_L)_{ii'}$
are the radial matrix elements of the vertex $g_L(r)$. Notice that  we deal with the channel with $I{=}0,\, S{=}0,\, L{=}0$. Therefore, the states $i,i'$ in (\ref{Vtild}) possess the same  SP angular momenta, $j_1{=}j_{1'},l_1{=}l_{1'};\,j_2{=}j_{2'},l_2{=}l_{2'}$.

In the valence subspace we consider, always there is   only one  state for each $(l,j)$ value.
Therefore, we need  only diagonal elements $\delta\Sigma_{\lambda \lambda}$. Explicit expression for the corresponding pole term is as follows \cite{DMD1,levels}:
\beq\begin{aligned}\delta\Sigma^{\rm pole}_{\lambda\lambda}(\epsilon) &= \\ {=} \sum_{\lambda'\,M} |\langle\lambda'|g_{LM}|\lambda\rangle|^2 &\left(\frac{n_{\lambda'}}{\eps{+}\omega_L{-} \eps_{\lambda'}}{+}\frac{1-n_{\lambda'}}{\eps{-}\omega_L {-}\eps_{\lambda'}}\right).\end{aligned}\label{dSig2}\eeq

In \cite{DMD1,DMD2}, the above equations were successfully applied to finding PC corrections to DMD values in magic nuclei. In semi-magic nuclei we deal, the PT solution (\ref{eps-Z-fac}) becomes regularly not valid because of the presence of the low-lying
$2^+$ state. For example, in the $^{204}$Pb nucleus we have $\omega_2{=}0.88\;$MeV and $\eps(3s_{1/2}){-}\eps(2d_{3/2}){=}0.79\;$MeV. As a result, catastrophically small energy denominators of ${\simeq} 0.1\;$MeV appear in Eq.~(\ref{Vind}) and in Eq.~(\ref{dSig2}) at $\eps{=}\eps_{\lambda}$. To find a direct solution of Eq.~(\ref{sp-eq}) in the valence space, instead of the PT one (\ref{eps-Z-fac}), is a key step to solve the problem.
It was made by us recently \cite{SPEnPB}. Here, we describe in short this method.

As it was mentioned above, only diagonal matrix elements of $\delta\Sigma_{\lambda \lambda}$ participate in equations in the valence subspace. In the result, Eq.~(\ref{sp-eq}) reduces as follows: \beq \eps-\eps_{\lambda} -\delta \Sigma^{\rm PC}_{\lambda \lambda}(\eps) =0,\label{sp-eq1}\eeq where $\delta \Sigma^{\rm PC}(\eps){=}\delta \Sigma^{\rm pole}(\eps){+}\delta \Sigma^{\rm tad}$, with obvious notation. The tadpole term does not depend on the energy, therefore the singular points of Eq.~(\ref{sp-eq1}) coincide with poles of Eq.~(\ref{dSig2}). They can be readily found from (\ref{dSig2}) in terms of $\eps_{\lambda}$ and $\omega_L$. It can be easily seen that the lhs of Eq.~(\ref{sp-eq1}) always changes sign between any couple of neighboring poles, therefore the corresponding solution $\eps_{\lambda}^i$ can be found with usual methods. In this notation, $\lambda$ is just the index for the initial SP state from which the state $|\lambda,i\rangle$ originated. The corresponding SP strength distribution factors ($S$-factors) are now determined with the energy derivative in the exact SP energy value: \beq S_{\lambda}^i =\left(1- \left(\frac {\partial} {\partial \eps} \delta \Sigma^{\rm PC}(\eps) \right)_{\eps=\eps_{\lambda}^{i}}\right)^{-1}. \label{S-fac}\eeq

\begin{figure}
\centerline {\includegraphics [width=9mm]{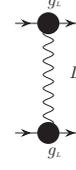}} \caption{Fig.~\ref{fig:PC2}. The $L$-phonon induced interaction.} \label{fig:PC2}
\end{figure}

In the result, the one-particle Green function in the subspace $S_0^{\rm v}$, which, without PC corrections, for each $\lambda$ contained only one pole, $G_{\lambda}(\eps) {=}
(\eps -\eps_{\lambda}\pm i\delta)^{-1}$, is now split into a sum of poles: \beq \tilde G_{\lambda} (\eps) = \sum_i G_{\lambda}^i(\eps),\;\;  G_{\lambda}^i (\eps)= \frac {S_{\lambda}^i} {\eps -\eps_{\lambda}^i\pm i\delta}\; .\label{Glam} \eeq

The correct scheme should involve, for the valence subspace, the insertion of the partial Green functions $G_{\lambda}^i$  instead of $G_{\lambda}$. As a  result, a total number of  the above relations strongly grows going from $\eps_{\lambda},Z_{\lambda}$ to $\eps_{\lambda}^i,S_{\lambda}^i$. Such approach is technically possible but it is rather cumbersome. We prefer to use an approximate approach suggested in \cite{SPEnPB} for finding the PC corrected SP energies. As the analysis shows, there are two different kinds of solutions of  (\ref{sp-eq1}). Their examples are shown in  Table~1. There are ``good'' SP states with a dominating term $|\lambda,i_0\rangle$ for which the following prescription similar to (\ref{eps-Z-fac}) looks natural: \beq \tilde \eps_{\lambda}={\eps}_{\lambda}^{i_0}; \;
Z_{\lambda} =S_{\lambda}^{i_0}. \label{i0} \eeq  It should be stressed that these relations  remind the PT solution (\ref{sp-eq1}) only in the form. Indeed, now ${\eps}_{\lambda}^{i_0}$ is one of the exact solutions of  (\ref{sp-eq1}). In addition, the $Z$-factor is determined now with the energy derivative (\ref{S-fac}) of the mass operator in this exact SP energy value.

There are also the cases of a strong spread where several terms $|\lambda,i\rangle$ possess comparable strengths $S_{\lambda}^i$. In such cases, the following generalization of Eq.~(\ref{i0}) was suggested in \cite{SPEnPB}: \beq  \tilde\eps_{\lambda}= \frac 1 {Z_{\lambda}}\sum_i
S_{\lambda}^i \eps_{\lambda}^i,\; Z_{\lambda} =\sum_i
S_{\lambda}^i.\label{spread2}\eeq According to \cite{SPEnPB}, all the states $|\lambda,i\rangle$ with comparably large strengths should be included into both the above sums.

\begin{table}[]
\begin{center}
\caption{Table~1. Examples of solutions of Eq.~(\ref{sp-eq1}) for protons in $^{204}$Pb.}
\begin{tabular}{|c|c|c|l|}

\hline

$\lambda$  & $i$ & $\eps_{\lambda}^i$ (MeV) &\hspace*{7mm} $S_{\lambda}^i$   \\

\hline

2$d_{5/2}$ &  1  &   -11.817 &  0.314 $\times 10^{-2}$ \\
           &  2  &   -11.150 &  0.139 \\
           &  3  &    -9.799 &  0.516 $\times 10^{-1}$ \\
           &  4  &    -8.580 &  0.312 \\
           &  5  &    -8.195 &  0.295 \\
           &  6  &    -7.404 &  0.171 \\
\hline

3$s_{1/2}$  &  1  &    -9.877  &  0.608 $\times 10^{-1}$ \\
            &  2  &    -8.536  &  0.604 $\times 10^{-1}$ \\
            &  3  &    -6.493  &  0.839 \\
\hline

\end{tabular}\label{tab1}
\end{center}
\end{table}

\begin{table}[]
\begin{center}
\caption{Table~2. PC corrected proton single-particle characteristics $\tilde \eps_\lambda$ (MeV) and $Z_\lambda$
of even Pb isotopes.}
\begin{tabular}{|c|c|c|c|c|}
\hline

\!\!Nucleus\!\!& $\lambda$ & $\eps_\lambda$ &$\tilde \eps_\lambda$ & $Z_\lambda$ \\
\hline

\!\!$^{200}$Pb& 1i$_{13/2}$
                         &   -0.26   &   -0.93   &   0.64   \\ 
          & 2f$_{7/2}$   &   -1.05   &   -1.35   &   0.83   \\ 
          & 1h$_{9/2}$   &   -2.33   &   -2.65   &   0.67   \\ 
          & 3s$_{1/2}$   &   -5.81   &   -5.32   &   0.77   \\ 
          & 2d$_{3/2}$   &   -6.67   &   -5.88   &   0.52   \\ 
          & 1h$_{11/2}$  &   -7.06   &   -6.39   &   0.72   \\ 
          & 2d$_{5/2}$   &   -7.88   &   -7.60   &   0.88   \\ 
          & 1g$_{7/2}$   &   -9.97   &   -9.89   &   0.91   \\ 

\hline

\!\!$^{202}$Pb& 1i$_{13/2}$
                         &   -0.74   &   -1.40   &   0.65  \\ 
          & 2f$_{7/2}$   &   -1.52   &   -1.81   &   0.83  \\ 
          & 1h$_{9/2}$   &   -2.86   &   -3.16   &   0.68  \\ 
          & 3s$_{1/2}$   &   -6.26   &   -5.75   &   0.77  \\ 
          & 2d$_{3/2}$   &   -7.09   &   -6.31   &   0.54  \\ 
          & 1h$_{11/2}$  &   -7.52   &   -6.87   &   0.73  \\ 
          & 2d$_{5/2}$   &   -8.34   &   -8.04   &   0.87  \\ 
          & 1g$_{7/2}$   &  -10.46   &   -10.38  &   0.91  \\ 

\hline

\!\!$^{204}$Pb& 1i$_{13/2}$
                         &   -1.21   &   -1.63   &   0.71  \\ 
          & 2f$_{7/2}$   &   -2.01   &   -2.24   &   0.87  \\ 
          & 1h$_{9/2}$   &   -3.36   &   -3.45   &   0.76  \\ 
          & 3s$_{1/2}$   &   -6.72   &   -6.49   &   0.84  \\ 
          & 2d$_{3/2}$   &   -7.51   &   -7.03   &   0.58  \\ 
          & 1h$_{11/2}$  &   -7.98   &   -7.51   &   0.81  \\ 
          & 2d$_{5/2}$   &   -8.80   &   -8.63   &   0.92  \\ 
          & 1g$_{7/2}$   &  -10.93   &   -10.49  &   0.54  \\ 

\hline

\!\!$^{206}$Pb& 1i$_{13/2}$
                         &   -1.67   &   -1.94   &   0.77  \\ 
          & 2f$_{7/2}$   &   -2.51   &   -2.81   &   0.82  \\ 
          & 1h$_{9/2}$   &   -3.82   &   -3.77   &   0.84  \\ 
          & 3s$_{1/2}$   &   -7.18   &   -7.10   &   0.89  \\ 
          & 2d$_{3/2}$   &   -7.91   &   -7.64   &   0.65  \\ 
          & 1h$_{11/2}$  &   -8.42   &   -8.05   &   0.88  \\ 
          & 2d$_{5/2}$   &   -9.28   &   -9.16   &   0.95  \\ 
          & 1g$_{7/2}$   &  -11.36   &   -11.12  &   0.90  \\ 
\hline
\end{tabular}\label{tab2}
\end{center}
\end{table}

\begin{table*} \caption{Table~3. Double mass differences $D_2$ (MeV) with and without account for PC effects and separate PC corrections to the $D_2$ values  in even Pb isotopes.\label{tab:delPCD2pm}}
\begin{center}
\begin{tabular}{|c|c|c|c|c|c|c|c|c|c|c|}
\hline
A&   & $D_2^{(0)}$  & $D_2(\gamma{=}0.06)$& $\delta D_2 (Z)$
  & $\delta D_2 ({\cal V}_{\rm ind} )$    &  $\delta D_2(\delta \eps)$ & $\delta D_2^{\rm PC}$ & $D_2^{\rm PC}$ &
   $D_2^{\rm PC}(\gamma{=}0.03)$   & $D_2^{\rm exp}$   \\
\hline

200   & $D_2^-$&  1.448& 0.923   & -0.897 &   1.807  &   0.892  & -0.208   &  1.240  &   1.166  &   1.0878(675)        \\
      & $D_2^+$& -2.099&-1.381   & 1.482  &  -1.603  &  -0.899  &  0.767   & -1.332  &  -1.219  &  -1.345(56)       \\

202   & $D_2^-$&   1.492&0.936  & -0.937 &   1.521  &  0.760   & -0.303   &  1.189  &  1.117   &   1.0856(324) \\
      & $D_2^+$&  -2.139&-1.396  & 1.500  &  -1.647  & -0.791   & 0.801    & -1.338  &  -1.225  &  -1.233(41)  \\

204   & $D_2^-$& 1.549& 0.956 & -0.875  &  1.311   &  0.886   &  -0.630   &  0.919   &  0.824   &   0.9333(44)    \\
      & $D_2^+$&-2.185&-1.415  & 1.373   & -1.287   &  0.780   &  0.939    &  -1.246  &  -1.110  &  -1.1694(152) \\

206   & $D_2^-$&  1.617& 0.981 & -0.794  &  0.290   &   0.510   & -0.613    &  1.004   &   0.856  &  0.8347(43)  \\
      & $D_2^+$& -2.236&-1.439 & -1.119  & -0.189   &  -0.369   & -0.918    & -1.318   &  -1.136  & -1.1461(78)  \\
\hline

 208  & $D_2^-$&   1.680&1.000  &  -0.824   &   -0.083  &   0.569  & -0.745  &  0.935  &  0.915  & 0.6271(312)  \\
      & $D_2^+$&  -2.286&-1.466 & 1.049     &  -0.167   &  -0.329  &  0.830  & -1.456  & -1.276  & -1.1844(52)  \\
\hline
$\langle\delta D_2\rangle_{\rm rms}$ &  &0.849 &   0.216  & - &- &- &- &0.167   &   0.087     &   -   \\

\hline
\end{tabular}
\label{tab3}
\end{center}
\end{table*}

In this approximate scheme, the exact Green function (\ref{Glam}) is changed with the approximate one,
\beq \tilde G_{\lambda} (\eps) =  \frac {Z_{\lambda}} {\eps -\tilde \eps_{\lambda} \pm
i\delta},\label{Glam_avr} \eeq where $\tilde \eps_{\lambda}$ and $Z_{\lambda}$  are taken  from
(\ref{i0}) or (\ref{spread2}), depending on the type of the solution we deal.

The final recipe to find DMD values in semi-magic nuclei we suggest is to use these non-perturbative
SP energies and $Z$-factors from Eqs.~(\ref{i0}) or (\ref{spread2}) in the set of equations
(\ref{eqchi0})--(\ref{Vind}), instead of the PT values used in  \cite{DMD1,DMD2} for magic nuclei.  In
this work, we test this method considering the same four even semi-magic $^{200-206}$Pb isotopes, as
in \cite{SPEnPB} where the method of direct solution of Eq.~(\ref{sp-eq1}) without any PT was
developed. Other technical details are also the same as in \cite{SPEnPB}, i.e. we use the DF3-a
version \cite{DF3-a} of the Fayans energy density functional \cite{Fay} to generate the
self-consistent basis $|\lambda\rangle$ and  take into account two $L$-phonons, $2^+_1$ and $3^-_1$.
Their characteristics may be found in \cite{SPEnPB}.

Table~2 contains the values of characteristics of the approximate PC corrected Green function
(\ref{Glam_avr}) we use. Some of them are different of those in the corresponding table in
\cite{SPEnPB}. The reason of that is in  different ways to choose the components 'i' in the sums of
(\ref{spread2}) for solutions with large spread. In \cite{SPEnPB}, we oriented to a procedure which is
used for finding the experimental SP energies when, in an odd nucleus under     consideration, the
excitations with the same $j^{\pi}$ are included in the sums of (\ref{spread2}) provided they possess
comparatively large spectroscopic factors $S^i(j^{\pi})$. This recipe is reasonable for theoretical
applications provided the exact Green function (\ref{Glam}) is integrated with a smooth energy
function. Now, this is not the case. Indeed, the use of two Green functions (\ref{Glam}) to find an
exact expression for the induced interaction instead of (\ref{Vind}) will result in a similar
expression with the  denominators $(\omega_L^2-(\eps_2^{i_2}-\eps_1^{i_1})^2 )$, with obvious
notation. In the case of the $2^+$ phonon, $\omega_2\simeq 1\;$MeV, this is rather sharp function of
two  energies in this expression, and contributions of the terms with smaller denominators are
enhanced. Therefore, in choosing the terms 'i' in Eq.~(\ref{spread2}) now we take into account the
``denominator factor'', in addition to the value of the spectroscopic factor.

Table~3 contains the results of the calculations of the DMD values with and without account for PC
effects. The initial DMD value denoted as $D_2^{(0)}$ is found on the base of the FP, i.e. from
Eqs.~(\ref{Vef1}) and (\ref{eqchi}) at $\gamma{=}0$. The next column contains similar quantity found
at  $\gamma{=}0.06$, which is the optimal value of this parameter found without account for PC effects
\cite{BCS50,Gnezd-1,Gnezd-2}. The next three columns present separate contributions of three different
PC effects under two others being switched off. For example, $\delta D_2 (Z)$ is the difference
between the $D_2$ value, found from Eqs.~(\ref{eqchi0}) and (\ref{Vtild}) at ${\cal V}_{\rm ind}{=}0$
and $\tilde\eps_{\lambda}{=}\eps_{\lambda}$, and the initial value of $D_2^{(0)}$. The next difference
$\delta D_2 ({\cal V}_{\rm ind} )$ is found according the same scheme, but now the induced interaction
${\cal V}_{\rm ind}$ in (\ref{Vtild}) is  taken into account at $Z_1{=}Z_2{=}1$. Finally, the quantity
$\delta D_2(\delta \eps)$ is found from (\ref{eqchi0}) when the difference of the SP energies
$\tilde\eps_{\lambda}$ from the initial values $\eps_{\lambda}$ is taken into account only. The
quantity  $\delta D_2^{\rm PC}{=} D_2^{\rm PC}{-}D_2^{(0)}$ shows the total PC effect. It should be
stressed that the total PC correction does not equal to the sum of the three separate ones as there is
some interference. For example, the induced interaction in (\ref{Vtild}) is multiplied with the
$Z$-factors. For completeness, we added the PT results for the magic $^{208}$Pb from \cite{DMD1,DMD2}.
As we see from the table, the corrections due to the $Z$-factor and due to the induced interaction
are, as a rule, very big and have opposite signs, the result being essentially smaller in absolute
value of each of them. Sometimes, the SP energy correction is also significant. The last two columns
preceding the experimental one contain the total PC corrected DMD values. The second of them includes
also the phenomenological addendum in Eq.~(\ref{Vef1}) with $\gamma{=}0.03$. This value is two times
less than the optimal one found previously without PC corrections. We see, that both the PC corrected
results for DMDs agree with experiment sufficiently well, especially the last of them. To estimate the
agreement with experimental data quantitatively, the rms differences between theoretical predictions
and data are given in the end of Table~3 for four versions of the theory: the pure FP calculation, the
result of the semi-microscopic model (\ref{Vef1}) with the value of $\gamma{=}0.06$ found previously
in calculations without PC corrections, and two results with the PC corrections, with $\gamma{=}0$ and
$\gamma{=}0.03$. The rms values of this differences are given in the last line of Table~3. One can see
that inclusion of the PC corrections makes agreement with experiment essentially better, especially in
a combination with a small phenomenological addendum of the semi-microscopic model
\cite{Pankrat-1,Pankrat-2} with $\gamma{=}0.03$.

To conclude, we developed for semi-magic nuclei a method of finding the PC corrections to the DMD
values in the approach starting from a free $NN$ potential. The main difference from the similar
problem for magic nuclei \cite{DMD1,DMD2} is that the PT used in magic nuclei for finding SP energies
and $Z$-factors is now unapplicable. Instead of this, we apply the method of the direct solution of
the Dyson equation, without any use of PT, developed by us recently \cite{SPEnPB}. The SP energies and
$Z$-factors, found in such a way, are now used in all expressions for the PC corrections under
consideration. Account for the PC corrections makes agreement of the DMD values with experiment
significantly better, especially in the version of the semi-microscopic model with the value of the
phenomenological parameter $\gamma{=}0.03$, which is two times less than the one in the approach
without PC corrections. As it was discussed when the semi-microscopic model was suggested
\cite{BCS50,Pankrat-1,Pankrat-2}, the phenomenological addendum proportional to the parameter $\gamma$
should take into account approximately three many-body effects changing the result of a simple FP
calculation. These are the difference of the effective mass of a nucleon inside a nucleus from a bare
one, the contribution from high-lying nuclear excitations as Giant Resonances, and finally, the PC
effects. However it is known \cite{BCS50}, that the first two effects possess opposite signs and
cancel each other significantly. In such a situation, the PC correction takes center stage. Our
calculation confirms this analysis. Indeed, the account for PC corrections diminishes the value of
$\gamma$, as a minimum, in two times. The analysis of a more wide base of data is necessary for a more
accurate estimate of the $\gamma$ value. In addition, the next refining of the calculation scheme is
desirable which includes the change of the approximate single-particle Green functions we use by the
exact representation of Eq.~(\ref{Glam}), where each single-particle pole is split into a sum of
several poles.

\vskip 0.3 cm This research was supported by the Russian Science Foundation, Grants Nos. 16-12-10155
and 16-12-10161. E.S. and S.P. thank the INFN, Seczione di Catania, for hospitality. Calculations were
partially carried out on the Computer Center of Kurchatov Institute.

{}

\end{document}